\documentclass{article}
\usepackage{amsmath,amsthm}
\theoremstyle{plain}
\newtheorem{thm}{Theorem}

\theoremstyle{remark}

\theoremstyle{definition}

\begin{document}

\title{Properties of the Spatial Sections of the Space-Time of a Rotating System.}
\author{Paschalis G. Paschali\footnote{email:pashalis@cytanet.com.cy}\\
Department of Mathematics and Statistics\\
University of Cyprus\\
Nicosia, P.O.Box 537\\
Cyprus\\
Georgios C. Chrysostomou\footnote{email: eng.cg@fit.ac.cy}\\
Frederick University, Pallouriotissa,
1036 Nicosia, Cyprus}
\date{\today}

\maketitle

\begin{abstract}
We study the symmetry group  of the geodesic equations of the 
spatial solutions of the space-time generated by  a noninertial rotating 
system of reference. It is a  seven dimensional Lie group, which is neither solvable nor nilpotent. 
The variational symmetries form a five dimensional solvable subgroup. 
Using the symplectic structure on the cotangent bundle we study the resulting 
Hamiltonian system, which is closely related to the geodesic flow on the spatial 
sections. We have also  studied some intrinsic and extrinsic geometrical 
properies of the spatial sections. 
\end{abstract} 

\subsubsection*{Keywords}

\section{Introduction}
\label{s:intro}
 In Classical Mechanics and prerelativistic science, space and time are separated and independent. 
 The time is absolute but the space is not. In the theory of Relativity both space and time are relative, 
 yet there is a  combination of them which is absolute. In this sense this theory introduces the concept of 
 space-time. This invariant combination can be expressed by the metric. In the Special Theory of Relativity 
 the space-time manifold is flat and its metric with respect to an inertial system of reference takes the well known 
 Lorentzian form  
\begin{equation}
ds^2 =-c^2dt^2+dx^2+dy^2+dz^2
\end{equation} 
In the General Theory of Relativity the presence of a gravitational field affects 
the geometry of space-time. Here the space-time manifold is curved. 
Actually the gravitational field is the curvature of the space-time and 
Einstein's equations relate the distribution of mass  and energy with the curvature 
of space-time. 

The cornerstone of the  general theory is the principle of equivalence, 
which states that every gravitational field is locally (but not globally) 
equivalent to a noninertial system of reference. If there is  no gravitational field 
but we decide to use a noninertial system of reference the expression of the metric 
will not be the above Lorentzian metric of Eq. (1), but still all the components   
of the Riemann tensor will be zero, since there is a global transformation from the inertial system 
to the noninertial that we are using. So with respect to a noninertial system of coordinates 
the space-time is still flat, but on the other hand the space itself may be curved. 
What we will study here are some geometrical properties of the spatial part of the 
space-time with respect to a rotating coordinate system with constant angular velocity $\omega$ 
around the $z$-axis. This is one of the simplest possible cases. Using similar techniques 
we can study more complicate problems resulting either from a more complicate system of reference, or 
because of the presence of a gravitational field. These problems can help us to understand 
better the principle of equivalence as well as the relation between gravity and the noninertial systems. 
We can also study the possibility of changes in the topology of the spatial part of the space-time.

Substituting $x$ and $y$ by $x\cos \omega t - y\sin \omega t$ and $x\sin \omega t + y\cos \omega t$ in Eq. (1) 
we obtain: 
\begin{equation}
ds^2 =\big [\omega^2 (x^2 +y^2)-c^2\big]dt^2+dx^2+dy^2+dz^2-2\omega ydxdt+2\omega xdydt
\end{equation}
The geometry of the space with respect to this rotating observer 
is not Euclidean since the relation between the circumference and 
the diameter of a circle is not any more $\pi$. Due to the Lorentz contraction 
along the circumference this is larger than $\pi$ by a factor of $(1-\omega^2 r^2 /c^2)^{-1/2}$. 
In cylindrical coordinates the above metric becomes: 
\begin{equation}
ds^2 =(\omega^2 r^2-c^2)dt^2+2\omega r^2d\phi dt+dz^2 +r^2 d\phi^2 +dr^2
\end{equation}

In this paper we have found the intrinsic and extrinsic curvatures 
of the spatial sections of the above space-time. They are both negative 
and blow up where the coordinate system we are using ceases to have physical meaning. 
We have also found the symmetry group of the geodesic equations and its variational subgroup. 
The first one is neither solvable nor nilpotent. On the other hand the variational subgroup is 
solvable. Using the Hamilton-Jacobi theory  for the Hamiltonian system which is induced 
on the cotangent bundle we have found the expression of the geodesic flow in terms of the affine time $t$. 

Account of the theory of Relativity can be found for example in the books of 
Misner, Thorne and Wheeler [11], Weinberg [16], O' Neill [13], Hawking and Ellis [7] and 
Landau and Lifschitz [8].

\section{The geometry of spatial sections.}

Let the metric of a space-time be given by: 
\begin{equation}
ds^2 =g_{\alpha \beta} dx^ \alpha dx^ \beta
\end{equation}
The Greek indices take the values $0,1,2,3$ and the Latin indices the values $1,2,3$. 
The index zero corresponds to the time of the event and $x^1, x^2, x^3$ determine its location in space. 
As a result we have: 
\begin{equation}
ds^2 =g_{00} dx^0dx^0 +2g_{0i}dx^0 dx^i +g_{ij}dx^i dx^j 
\end{equation}
to study the spatial sections of the space-time manifold we must consider the set of all 
events that are simultaneous to a given event. Using the null geodesics we can prove 
that two nearby events are simultaneous if they satisfy the condition $(dx)_0 =0$. As 
a result the metric on the spatial sections in contravariant form is  
\begin{equation}
dl^2=g^{ij}dx_i dx_j \nonumber
\end{equation}
and in covariant form it becomes 
\begin{equation}
dl^2 = \Big( g_{ij} -\frac{g_{0i}g_{0j}}{g_{00}}  \Big)dx^i dx^j
\end{equation} 
Applying this to Eqs. (2) and (3) we get the metric on the spatial sections 
of a noninertial rotating system in cartesian coordinates in the form: 
\begin{equation} 
dl^2 =(1+Ay^2)dx^2-2Axydxdy+(1+Ax^2)dy^2+dz^2
\end{equation}
In cylindrical coordinates we have 
\begin{equation} 
dl^2 =dr^2 +\frac {r^2}{1-\omega^2 r^2 /c^2}d  \phi^2 +dz^2
\end{equation}
In Eq. (7) the factor $A$ is given by: 
\begin{equation} 
A=\frac{\omega^2}{c^2 -\omega^2 (x^2 +y^2)} \nonumber
\end{equation}
The metric in Eq. (8) is static so it can be used to define and to calculate 
finite distances and the result will not depend on the cosmic curve connecting the two points. 
On the other hand we must bear in mind that the concept of simultaneous events which 
is the foundation of the positive definite spatial metric of Eq. (6) is only a local one 
and it may not be possible to define simultaneous events in a consistent way over the 
wlole manifold since $g_{0i} \ne 0$
These arguments indicate that the concept of a  spatial slice is maybe 
more complicate than the concept of a hypersurface for example and here may appear 
some questions of topological character. 

We turn now to some geometrical properties of the metric (8). The $z$-axis is a geodesic 
parametrized by arclength that is the same with the coordinate $z$. 
This is not true for the $x, y, r$ or $\phi$ coordinates. This is a space of negative curvature. 
The Riemann tensor has only 6 functionally independent components that are all zero except 
$R_{1212}$ , which is given by: 
\begin{equation} 
R_{1212} =-\frac{3c^4\omega^2 r^2}{(c^2 -\omega^2 r^2)^3}
\end{equation} 
Here we use the notation $x^1 =r, x^2 =\phi$ and $x^3 =z$. We can easily 
calculate the scalar curvature and get the following result: 
\begin{equation} 
R =g^{ki}g^{lj}R_{klij}=-\frac{6c^2\omega^2 }{(c^2 -\omega^2 r^2)^2}
\end{equation} 
This is negative and blows up when $r=c/\omega$. This is exactly where the coordinates 
of the rotating system cease to have meaning since beyond that distance we have speeds 
higher than the speed of light in vacuum. Using the metric (8) we can get  for the element 
of the arclength along the circle with both $z, r$ constant the expression:  
\begin{equation} 
dl=\frac{cr}{\sqrt{c^2 -\omega^2 r^2}}d\phi \nonumber
\end{equation}
from where we get that the ratio of the circumference to the diameter is greater than $\pi$ 
by the Lorentz factor. 

The intrinsic geometry does not give any information about how the spatial slices 
are impended in the four dimensional space-time. Geometrically the impending can 
be found by studying the unit normal vector on the spatial sections and is somehow 
represented by the extrinsic curvature, which is a symmetric tensor parallel to the 
spatial slices. The extrinsic curvature is closely related to the Weingarten map, 
which is the Jacobian of the Gauss map and can be calculated as the covariant derivative 
of the unit normal along the spatial slices. The metric (4) in the canonical $1+3$ 
language takes the form: 
\begin{equation}
\begin{split}
ds^2 & = g_{ij}(dx^i +N^i dt)(dx^j +N^j dt)-N^2 dt^2 = (g_{ij} N^i N^j -N^2 )dt^2\\
&+g_{ij}N^j dx^i dt+g_{ij}N^i dx^j dt +g_{ij}dx^i dx^j\\ 
\end{split}
\end{equation} 
where $N^i$ are the shift functions and $N$ is the lapse function. Comparing this expression with Eq. (8) 
we get the shift and lapse functions of our space-time: 
\begin{equation}
N^r =N^z =0
\end{equation}
\begin{equation}
N^\phi =\omega
\end{equation}
\begin{equation}
N=c
\end{equation}
So the contravariant normal vector connecting the spatial sections that correspond to time 
$t$ and $t+dt$ is 
\begin{equation}
(dt, 0, -\omega dt, 0) \nonumber
\end{equation}
and its proper length is  $cdt$ as we should expect. So the unit contravariant vector is 
\begin{equation}
\eta^\alpha =\Big( \frac{1}{c}, 0, -\frac{\omega}{c}, 0 \Big) 
\end{equation}
In covariant form is 
\begin{equation}
\eta_\alpha =( c, 0, 0, 0) 
\end{equation} 
Since the metric $g_{ij}$ is static, the tensor of the extrinsic curvature can be 
calculated using the following relation:
\begin{equation}
K_{ij} =\frac{1}{2N}\Big( N_{j;i} +N_{i;j} \Big) 
\end{equation} 
where the semicolon represents covariant differentiation with respect to the 
metric of the spatial sections. We can easily calculate the quantities $N_r, N_\phi$ and $N_z$ 
and we obtain: 
\begin{equation}
N_r =N_z =0
\end{equation} 
\begin{equation}
N_\phi =\frac{\omega c^2 r^2}{c^2 -\omega^2 r^2}
\end{equation}  
Using these relations in (17) we can prove that all the components of the extrinsic curvature 
are zero except the component $K_{r\phi}$, which is given by: 
\begin{equation}
K_{r\phi} =-\frac{\omega c^3 r}{(c^2 -\omega^2 r^2)^3} 
\end{equation} 
and it has again a singularity at the positions $r=c/\omega$. The Riemannian 
tensor $R_{ijkl}$ and the above extrinsic curvature tensor are related by the 
Gauss-Codazzi equations.   

\section 
{The symmetry group of the geodesics.} 
Here we shall study the Lie point symmetries  and the variational symmetries  
of the geodesic equations. A symmetry group of a system of differential equations 
is a group acting on the space of independent and depandent variables in such a way that solutions 
are mapped into other solutions. A local group of transformations is a variational symmetry if it 
necessarily leaves the variational integral invariant. Every variational group is 
a symmetry group of the corresponding Euler-Lagrange equations but the opposite is not true. 
The symmetry approach to differential equations can be found, for example, in the books of Olver [12], 
Bluman and Cole [3], Bluman and Kumei [4], Fushchich and Nikitin [6] and Obsiannikov [14]. 
In our problem the geodesics satisfy the following equations: 
\begin{equation}
\frac{d^2 x^l}{ds^2}+\Gamma_{jk}^l \frac{dx^j}{ds} \frac{dx^k}{ds} =0
\end{equation}  
where $\Gamma_{jk}^l$ are the Christoffel symbols and $s$ is the length of the geodesics, 
which we use to parametrize them. Here we do not have to worry about null geodesics since the metric (8) 
on the spatial sections is positive definite. Because of the form of this metric the Christoffel 
symbols are given by the following relations: 
\begin{equation}
\Gamma _{jk}^l =0
\end{equation}
\begin{equation}
\Gamma _{ii}^j =-\frac{1}{2g_{jj}} \frac{\partial g_{ii}}{\partial x^j}
\end{equation}
\begin{equation}
\Gamma _{ij}^i =\frac{1}{2} \frac{\partial \ln g_{ii}}{\partial x^j}
\end{equation}
\begin{equation}
\Gamma _{ii}^i =\frac{1}{2} \frac{\partial \ln g_{ii}}{\partial x^i}
\end{equation}
where the indices $i, j, k$ are all different. Using these in the metric (8) we get 
the differential equations of the geodesics in the form: 
\begin{equation}
\frac{d^2 r}{ds^2} - \frac{c^4 r}{(c^2 -\omega^2 r^2)^2} \Big ( \frac{d\phi}{ds} \Big)^2 =0 
\end{equation}
\begin{equation}
\frac{d^2 \phi}{ds^2} + \frac{2c^2}{r(c^2 -\omega^2 r^2)} \frac{d\phi}{ds} \frac {dr}{ds} =0 
\end{equation}
\begin{equation}
\frac{d^2 z}{dr^2} =0 
\end{equation}
We can also find the geodesics by considering the partial differential equation \begin{equation}
g^{ij} \frac{\partial w}{\partial x^i} \frac{\partial w}{\partial x^j} = 
\Big ( \frac{\partial w}{\partial r} \Big )^2 
+\frac{c^2 - \omega^2 r^2}{c^2 r^2} \Big ( \frac{\partial w}{\partial \phi} \Big )^2 
+ \Big ( \frac{\partial w}{\partial z}\Big )=1
\end{equation}
A complete solution of this equation has the form 
\begin{equation}
w(r, \phi , z, a_1 , a_2 ) =k \nonumber
\end{equation}
where $k$ is a constant. The equations of the geodesics in parametric form take the form: 
\begin{equation}
\frac {\partial w}{\partial {a_1}} = b_1 \nonumber
\end{equation}
\begin{equation}
\frac {\partial w}{\partial {a_2}} = b_2 \nonumber
\end{equation}
where $b_1$ and $b_2$ are arbitrary constants. 

Using the method of Lie groups we shall study the symmetry group of the geodesics. 
A vector field $G$ acting on the space of independent and dependent variables 
is a symmetry of the equations (26)-(28) if and only if it satisfies the relations 
\begin{equation}
pr^{(2)} G(Eq26)= pr^{(2)} G(Eq27)=  pr^{(2)} G(Eq28)= 0    \nonumber
\end{equation}
where $pr^{(2)} G$ is the second prolongation or extension of the vector field $G$. If 
\begin{equation}
G=\Sigma \frac{\partial}{\partial s} + R \frac{\partial}{\partial r} + \Phi \frac{\partial}{\partial \phi} 
+ Z \frac{\partial}{\partial z} 
\end{equation}
then 
\begin{equation} 
\begin{split}
pr^{(2)} G= & G +(\dot{R}-\dot{\Sigma}\dot{r})\frac{\partial}{\partial \dot{r}} +  
(\dot{\Phi}-\dot{\Sigma}\dot{\phi})\frac{\partial}{\partial \dot{\phi}} 
+ (\dot{Z}-\dot{\Sigma}\dot{z})\frac{\partial}{\partial \dot{z}}\\ 
&+ (\ddot{R}-\ddot{\Sigma}\dot{r}-2 \dot{\Sigma}\ddot{r})\frac{\partial}{\partial \ddot{r}}
+ (\ddot{\Phi}-\ddot{\Sigma}\dot{\phi}  
-2 \dot{\Sigma}\ddot{\phi})\frac{\partial}{\partial \ddot{\phi}} \\ 
&+(\ddot{Z}-\ddot{\Sigma}\dot{z}-2 \dot{\Sigma}\ddot{z})\frac{\partial}{\partial \ddot{z}}\\
\end{split}
\end{equation}
If we apply this to the geodesic equations (26)-(27) we get the following conditions: 
\begin{equation}
\begin{split}
\ddot{R}-\ddot{\Sigma}\dot{r}-&2\dot{\Sigma}\ddot{r} -\frac{2c^4 r}{(c^2 -\omega^2 r^2)^2}\dot{\phi}
(\dot{\Phi} -\dot{\Sigma} \dot{\phi})\\
&-\frac{c^4(c^2 +3\omega^2 r^2)}{(c^2 -\omega^2 r^2)^3}R \dot{\phi}^2 =0\\
\end{split}
\end{equation}
\begin{equation}
\begin{split}
&\ddot{\Phi}-\ddot{\Sigma}\dot{\phi}-2\dot{\Sigma}\ddot{\phi} +\frac{2c^2}{r(c^2 -\omega^2 r^2)}\dot{\phi}
(\dot{R} -\dot{\Sigma} \dot{r})\\
&+\frac{2c^2}{r(c^2 -\omega^2 r^2)}(\dot{\Phi}-\dot{\Sigma}\dot{\phi})\dot{r} 
-\frac{2c^2(c^2 -3\omega^2 r^2)}{r^2(c^2 -\omega^2 r^2)^2}R\dot{\phi}\dot{r}=0\\
\end{split}
\end{equation}
\begin{equation}
\ddot{Z}-\ddot{\Sigma}\dot{z}-2\dot{\Sigma}\ddot{z} = 0
\end{equation}
Expanding Eq. (34) and using Eqs. (26)-(28) we obtain: 
\begin{equation} 
\begin{split} 
&Z_{ss} + 2Z_{sr}\dot{r} +2Z_{s\phi}\dot{\phi} + 2Z_{sz}\dot{z}+ 2Z_{r\phi}\dot{r}\dot{\phi} 
+ 2Z_{rz}\dot{r}\dot{z} + 2Z_{\phi z}\dot{\phi}\dot{z}\\
&+Z_r a(r) \dot{\phi}^2 - Z_\phi b(r)\dot{\phi}\dot{r} +Z_{rr}\dot{r}^2 + Z_{\phi \phi}\dot{\phi}^2 
+Z_{zz}\dot{z}^2-\Sigma_{ss}\dot{z}\\
&-2\Sigma_{rs}\dot{z}\dot{r}-2\Sigma_{s\phi}\dot{z}\dot{\phi}-2\Sigma_{sz}\dot{z}^2 -2\Sigma_{r\phi}\dot{z}\dot{r}\dot{\phi} 
-2\Sigma_{rz}\dot{z}^2 \dot{r} -2\Sigma_{\phi z}\dot{\phi}\dot{z}^2\\
&-\Sigma_r a(r)\dot{z} \dot{\phi}^2 + \Sigma_\phi b(r) \dot{z} \dot{\phi}\dot{r} 
-\Sigma_{rr}\dot{z}\dot{r}^2 - \Sigma_{\phi \phi}\dot{z}\dot{\phi}^2 - 
\Sigma_{zz}\dot{z}^3 =0 \\
\end{split} 
\end{equation}
The method we used is similar with the one in [9] and [15]. Condition (35) after a long 
calculation gives the following for $\Sigma$ and $Z$: 
\begin{equation} 
\Sigma = \frac{1}{2}c_1s^2+c_2s+c_3+z(c_5s+c_6)
\end{equation} 
\begin{equation} 
Z = c_5z^2+c_8z+c_9+s\Big( \frac{1}{2}c_1z+c_7 \Big)
\end{equation} 
where all the
$c_i$ are arbitrary constants. Using the results above in Eqs. (32) and (33) after a long but 
straightforward calculation we obtain the following final result: 
\begin{equation}
\Sigma=c_2s+c_3z+c_4
\end{equation}
\begin{equation}
R=0
\end{equation}
\begin{equation}
\Phi =c_1
\end{equation}
\begin{equation}
Z=c_5s+c_6z+c_7
\end{equation}
So the symmetries of the geodesic equations form a seven dimensional group 
with generators 
\begin{equation} 
\vec{x_1}=\partial_\phi,  \vec{x_2}=s\partial_s, \vec{x_3}=z\partial_s, \vec{x_4}=\partial_s, 
\vec{x_5}=s\partial_z, \vec{x_6}=z\partial_z, \vec{x_7}=\partial_z \nonumber
\end{equation}  
The multiplication table 
for this Lie group is given by the relations: 
\begin{equation}
[\vec{x_2},\vec{x_3}]=-\vec{x_3}, [\vec{x_2},\vec{x_4}]=-\vec{x_4}, [\vec{x_5},\vec{x_2}]=-\vec{x_5}, 
\nonumber 
\end{equation} 
\begin{equation} 
[\vec{x_3},\vec{x_5}]=\vec{x_6}-\vec{x_2}, [\vec{x_3},\vec{x_6}]=-\vec{x_3}, [\vec{x_3},\vec{x_7}]=-\vec{x_4},
\nonumber 
\end{equation} 
\begin{equation} 
[\vec{x_4},\vec{x_5}]=\vec{x_7}, [\vec{x_5},\vec{x_6}]=-\vec{x_5}, [\vec{x_6},\vec{x_7}]=-\vec{x_7} \nonumber
\end{equation} 
The subroup $g=[G,G]$ is five dimensional and is generated by the vectors $\vec{x_2}-\vec{x_6}$, $\vec{x_3}$, 
$\vec{x_4}$, $\vec{x_5}$ and $\vec{x_7}$. \\
If we calculate the group $[g,g]$ we will see that this is the same as $g$. So the derived series of ideals 
terminates. So the symmetry group of the geodesics is neither nilpotent, nor solvable, in contrary to 
systems like  the Mawell-Bloch [5] or the Hilbert-Einstein equations in Cosmology [15]. 

Using Noether's theorems we can check which of these symmetries are variational and also we can 
find the corresponding  conserved quantities. A vector field $\vec{X}$ is a variational symmetry 
of a variational problem with Lagrangian $L$ if and only if it satisfies the following condition: 
\begin{equation}
pr^{(1)} \vec{X}+L\vec{\nabla}\cdot\vec{\xi}=0
\end{equation}
where 
\begin{equation}
\vec{X}=\sum \limits_{i=1}^n \lambda^i \frac{\partial}{\partial t^i} + 
\sum \limits_{j=1}^q \psi^j \frac{\partial}{\partial u^j}
\end{equation}
and 
\begin{equation}
\vec{\xi}=\sum \limits_{i=1}^n \lambda^i \frac{\partial}{\partial t^i} 
\end{equation}
Every variational symmetry $\vec{X}$ generates a conservation law, 
which can be written in the form: 
\begin{equation}
\vec{\nabla}\cdot \vec{P}=\sum \limits_{j=1}^q Q_j \cdot E_j (L)
\end{equation}
where $E_j(L)$ are the Euler-Lagrange operators acting on the Lagrangian and 
$Q_j$ are the characteristics of $\vec{X}$. In our case $L$ is given by Eq. (8). 
Using the conditions above we can prove that only the infinitesimal generators $\vec{X_i}$ with 
$i=1,2,3,4$ and $7$ are variational symmetries. They form the Lie subgroup $g_v$ of the 
full symmetry group. We can easily check that 
\begin{equation}
[g_v , g_v]=[g_v ,[g_v , g_v]]=\{ \vec{X_3} , \vec{X_4 } \} \nonumber
\end{equation}
and since $[\vec{X_3} , \vec{X_4 }]=0$ we infer that the group $g_v$ is solvable but not nilpotent. 
So we have the following theorem:
\begin{thm} 
\label{thm} 
The Group of symmetries of the geodesic equations 
of the spatial sections of the four dimensional space-time with respect to a 
uniformly rotating noninertial system of reference is a seven dimensional Lie group,  which is 
neither nilpotent, nor solvable. on the oher hand the rotational symmetries 
form a five dimensional subgroup, which  is solvable but not nilpotent. 
\end{thm}

The equations of Killing, which can be written in the form  
\begin{equation} 
\zeta_{j;i}+\zeta_{i;j}=0 
\end{equation}
determine the infinitesimal generators of the isometry group of a space. 
Here the semicolon represents covariant differentiation with respect 
to the metric(8). If in Eqs. (38)-(41) we set $\Sigma =c_5 =0$ we get the vector 
\begin{equation} 
c_1 \frac{\partial}{\partial \phi} + (c_6 z+c_7) \frac{\partial}{\partial z} \nonumber 
\end{equation} 
which acts only on the coordinates $r, \phi$ and $z$ of the spatial sections. 
This is of course a local symmetry of the geodesic equations so it preserves 
the geodesics. If we set $c_6=c_7=0$ we get the vector 
\begin{equation} 
\frac{\partial}{\partial \phi}  \nonumber 
\end{equation}
which not only preserves the geodesics, but is also a Killing vector. 
A second Killing vector can be found if we set $c_6=c_1=0$ and it is given by 
\begin{equation} 
\frac{\partial}{\partial z}  \nonumber 
\end{equation}
Finally for $c_1=c_6=0$ we get the vector 
\begin{equation} 
z\frac{\partial}{\partial z}  \nonumber 
\end{equation}
which preserves the geodesics but it is not a Killing vector because 
it does not satisfy 
\begin{equation} 
\frac{\partial \xi_z}{\partial z}=0  \nonumber 
\end{equation}
which is one of the Killing's equations. Taking into account the fact that every Killing vector preserves 
the geodesics we infer that the above arguments indicate that the spatial sections 
have only two Killing vectors, which are the ones shown above.

\section 
{The geodesics from the Hamiltonian point of view} 
Here we shall study the geodesics using the canonical symplectic structure 
on the cotangent bundle of the spatial sections [1], [2]. 

If $M$ is a Riemannian manifold with metric $g_{ij}$ we consider the Hamiltonian 
$H$ with local expression 
\begin{equation}
H(x^i ,p_i)=\frac{1}{2}g^{ij} (x^k)p_i p_j \nonumber
\end{equation}
where $x^i , p_i$ are canonical coordinates of the contangent bundle $T*M$. \\ 
We can easily prove that the projections of the inertial curves of $H$ onto $M$ 
are the geodesics of $g_{ij}$ parameterized by their affine parameter $t$. 
Hamilton's equations can be written in the form: 
\begin{equation}
\dot{x}^i =g^{il} p_l
\end{equation}
\begin{equation} 
\dot{p_i}=-\frac{1}{2}\frac{\partial g^{kl}}{\partial x^i}p_k p_l
\end{equation}
Inverting Eq. (47) we get 
\begin{equation} 
p_l =g_{il} \dot{x^i}
\end{equation}  
Differentiating Eq. (47) and using Eqs. (47) and (49) we can find 
\begin{equation} 
\ddot{x^i}=-\frac{1}{2}g^{il} \frac{\partial g^{mn}}{\partial x_l}p_mp_n +
\frac{\partial g^{il}}{\partial x^k}g^{km}p_mp_l
\end{equation} 
Using Eq. (49) and after some algebra we end up with: 
\begin{equation} 
\ddot{x^i}=g^{il} \Big[ \frac{1}{2} \frac{\partial g_{jk}}{\partial x^l} - \frac{\partial g_{lj}}{\partial x^k}\dot{x^k}\dot{x^j}\Big]
\end{equation} 
which is actually 
\begin{equation} 
\ddot{x^i}=\frac{1}{2}g^{il} \Big[ \frac{\partial g_{jk}}{\partial x^l} - 
\frac{\partial g_{lj}}{\partial x^k}-\frac{\partial g_{lk}}{\partial x^j}\Big]\dot{x^k}\dot{x^j}
\end{equation} 
Taking into account the definiotion of Christoffel symbols it becomes clear 
that here we have the geodesic Eqs.(21). \\
Using the metric (8) we get the Hamiltonian in the form: 
\begin{equation} 
H(x^i,p_i)=\frac{1}{2}p_r^2+\frac{1}{2}\frac{c^2 -\omega^2 r^2}{c^2 r^2}p_\phi ^2 +\frac{1}{2}p_z^2
\end{equation} 
We can easily prove that this Hamiltonian is separable since it satisfies 
the Levi-Civita conditions [10]. Thus we can solve this system using 
the Hamiltin-Jacobi method. Hamilton's characteristic function takes the form 
\begin{equation}
W=W_1 (r)+p_\phi \phi +p_z z
\end{equation}
and the Hamilton-Jacobi equation becomes 
\begin{equation}
\frac{1}{2}\Big(\frac{dW_1}{dr}\Big)^2 +\frac{1}{2}\frac{c^2 -\omega^2 r^2}{c^2 r^2}p_\phi ^2 +\frac{1}{2}p_z^2=E
\end{equation}
where $p_\phi, p_z$ and $E$ are constants. Integrating we get $W$ 
\begin{equation}
W=\frac{1}{c}\int \frac{1}{r}\sqrt{(2Ec^2 -c^2 p_z^2 +\omega^2 p_\phi ^2 )r^2-c^2 p_\phi ^2}dr 
+p_\phi \phi +p_z z
\end{equation}
At this stage we can write down the solution in the form 
\begin{equation}
t+c_1 =\frac{\partial W}{\partial E}
\end{equation}
\begin{equation}
c_2 =\frac{\partial W}{\partial p_\phi}
\end{equation}
\begin{equation}
c_3 =\frac{\partial W}{\partial p_z}
\end{equation}
where $t$ is the affine parameter of the geodesics and $c_1, c_2$ and $c_3$ are 
arbitrary constants. Using Eq. (55) and after some integrations we end up with the following results: 
\begin{equation}
t+c_1 =\frac{c}{A}\sqrt{Ar^2-B}
\end{equation}
\begin{equation}
c_2 =\phi +\frac{\omega^2 p_\phi}{c}\frac{\sqrt{Ar^2-B}}{A}+tan^{-1} \Big[ \frac{cp_\phi}{\sqrt{Ar^2-B}} \Big]
\end{equation}
\begin{equation}
c_3 =z-\frac{cp_z}{A}\sqrt{Ar^2-B}
\end{equation}
where $A$ and $B$ are given by: 
\begin{equation}
A=2Ec^2-c^2p_z^2+\omega^2 p_\phi ^2
\end{equation}
\begin{equation}
B=c^2p_\phi ^2
\end{equation}
In the Maxwell-Bloch system we found a  symplectic realization [5] and using it we have found 
the action angle variables and then we integrated the system using elliptic functions. 
In contrary as we have seen the above system can be integrated using elementary functions.

\end{document}